\documentclass[12pt]{revtex4}
\usepackage{amsmath,amssymb,graphicx}

\def\al{\alpha}

\def\ep{\epsilon}

\def\ka{\kappa}
\def\la{\lambda}

\def\si{\sigma}
\def\vs{\varsigma}

\def\om{\omega}

\def\mn{{\mu\nu}}

\def\prt{\partial}
\def\cl{{\mathcal L}}
\def\vev#1{\langle {#1}\rangle}

\def\fr#1#2{{{#1} \over {#2}}}
\def\Frac#1#2{{\textstyle{{#1}\over {#2}}}}
\def\half{{\textstyle{1\over 2}}}
\def\lsim{\mathrel{\rlap{\lower4pt\hbox{\hskip1pt$\sim$}}
    \raise1pt\hbox{$<$}}}
\def\gsim{\mathrel{\rlap{\lower4pt\hbox{\hskip1pt$\sim$}}
    \raise1pt\hbox{$>$}}}
\def\sqr#1#2{{\vcenter{\vbox{\hrule height.#2pt
         \hbox{\vrule width.#2pt height#1pt \kern#1pt
         \vrule width.#2pt}
         \hrule height.#2pt}}}}
\newcommand{\beq}{\begin{equation}}
\newcommand{\eeq}{\end{equation}}
\newcommand{\bea}{\begin{eqnarray}}
\newcommand{\eea}{\end{eqnarray}}
\newcommand{\rf}[1]{(\ref{#1})}

\def\kf{\hat k_F}
\def\kaf{\hat k_{AF}}
\def\kfd#1{k_{F}^{(#1)}}
\def\kafd#1{k_{AF}^{(#1)}}
\def\klm#1#2#3{k^{(#1)}_{(#2)#3}}
\def\sylm#1#2{\phantom{}_{#1}Y_{#2}}
\def\mbf#1{\mbox{\boldmath$#1$}}

\def\kE{\klm{d}{E}{lm}}
\def\kB{\klm{d}{B}{lm}}
\def\kV{\klm{d}{V}{lm}}
\def\kVdlm#1#2{\klm{#1}{V}{#2}}

\begin{document}

\title{Lorentz Violation, Electrodynamics, 
  and the Cosmic Microwave Background\footnote{
    Presented at the Fourth Meeting on CPT and Lorentz
    Symmetry, Bloomington, Indiana, August, 2007.}}

\author{Matthew Mewes}
\address{Physics Department, Marquette University, Milwaukee, WI 53201}

\begin{abstract}
  Vacuum birefringence is a signature of 
  Lorentz-symmetry violation.
  Here we report on a recent search for
  birefringence in the cosmic
  microwave background.
  Polarization data is used
  to place constraints on
  certain forms of Lorentz violation.
\end{abstract}

\maketitle

\section{Introduction}

The properties of light have proved 
to be a valuable testing ground for
special relativity for more than a century.
Contemporary experiments are motivated in part
by a possible breakdown of special relativity
with origins in Planck-scale physics
\cite{kost,cpt04,ck}.
These experiments include modern versions of the
classic Michelson-Morley and
Kennedy-Thorndike experiments
that use highly stable resonant cavities
to search for violations of
rotation and boost symmetries
\cite{cavities}.
However, the highest sensitivities
to relativity violations
in electrodynamics are found in
searches for vacuum birefringence in 
light from very distant sources
\cite{cfj,km1,km2,km3}.
Birefringence studies take advantage of the
extremely long baselines that allow the miniscule
effects of a Lorentz violation 
to accumulate to (potentially)
detectable levels over the billions of years
it takes for the light to reach Earth.
The cosmic microwave background (CMB)
is the oldest light available to observation
and therefore provides an excellent
source for birefringence searches.
Here we summarize a recent search for
signals of Lorentz violation
using CMB polarimetry
\cite{km3}.

General Lorentz violation is
described by a framework known
as the Standard Model Extension (SME)
\cite{ck}.
The SME provides the theoretical
backbone for studies in a number of areas
\cite{cpt04},
including photons
\cite{cavities,km1,km2,km3,photons}.
Most tests of Lorentz violation
focus on the minimal SME,
which assumes usual gauge invariance
and energy-momentum conservation
and restricts attention to superficially
renormalizable operators.
Operators of dimension $d\leq4$
are of renormalizable dimension
and are included the minimal SME.
Two types of Lorentz-violating operators
appear in the minimal SME,
$CPT$-odd operators with coefficients
$(k_{AF})_\ka$ and $CPT$-even operators
with coefficients $(k_F)^{\ka\la\mu\nu}$.

In this work,
we also consider non-minimal
higher-dimensional operators
in the photon sector with $d>4$.
In general there are an infinite
number of possible operators
that emerge when we relax the 
renormalizable condition.
These operators are
phenomenologically and theoretically 
relevant in that they help
provide a connection to the
underlying Planck-scale physics.
They also add a number of new and
interesting signals for Lorentz
violation that may be tested experimentally.

\section{Theory}

General Lorentz-violating
electrodynamics is given
by a lagrangian that takes the
same basic form as the minimal-SME
photon sector
\cite{km3}:
\beq
\cl = -\fr14 F_\mn F^\mn 
+ \fr12\ep^{\ka\la\mu\nu} A_\la (\kaf)_\ka F^{\mu\nu}
-\fr14 F_{\ka\la} (\kf)^{\ka\la\mu\nu} F_{\mu\nu} \ .
\eeq
We assume a linear theory
and impose the usual U(1) gauge invariance.
The key difference between this theory
and the minimal-SME photon sector 
is that here the $\kaf$ and $\kf$ coefficients
are differential operators.
The effects of these operators
mimic the effects of a permeable medium
whose activity depends on the
photon energy and momentum.
This introduces a plethora of new effects not
found in either the conventional 
Lorentz-conserving case or the minimal SME.
These include drastically different frequency
dependences and direction-dependent propagation
of light.

Expanding the $\kaf$ and $\kf$ operators
in the 4-momentum $p_\mu =  i\prt_\mu$
leads to the expressions
\begin{align}
  (\kaf)_\ka&=\sum {(\kafd{d})_\ka}^{\al_1\ldots\al_{(d-3)}}
  \prt_{\al_1}\ldots\prt_{\al_{(d-3)}} \ , \\
  \allowbreak
  (\kf)^{\ka\la\mu\nu}&=\sum (\kfd{d})^{\ka\la\mu\nu\al_1\ldots\al_{(d-4)}}
  \prt_{\al_1}\ldots\prt_{\al_{(d-4)}} \ .
\end{align}
The coefficients for Lorentz violation
associated with the dimension $d$ operators
are now given by 
${(\kafd{d})_\ka}^{\al_1\ldots\al_{(d-3)}}$ and
$(\kfd{d})^{\ka\la\mu\nu\al_1\ldots\al_{(d-4)}}$.
The $\kaf$ expression contains all
$CPT$-breaking effects, and the sum
is restricted to odd-dimensional operators,
$d=$ odd.
The $\kf$ coefficients control all $CPT$-even
violations and have $d=$ even.
Imposing gauge invariance places
various constraints on these coefficients.
For $\kafd{d}$ coefficients, any trace of the 
that involves the first index vanishes identically.
For $\kfd{d}$, the antisymmetrization on
any three indices vanishes.
Standard group theory techniques allow a
counting of the independent coefficients
for Lorentz violations
\cite{km4}.
For a given dimension $d$,
we find $\half(d+1)(d-1)(d-2)$
independent $\kafd{d}$ coefficients
in the $CPT$-odd case and $(d+1)d(d-3)$ 
independent $\kfd{d}$ coefficients
in the $CPT$-even case.

For studies of Lorentz-violation
induced birefringence,
certain linear combinations of these
general coefficients are important.
They result from a spherical-harmonic
expansion of plane-waves propagating
in the vacuum.
This plane-wave expansion is 
best characterized using the language
of Stokes parameters.
We begin by defining 
a Stokes vector $\mbf s = (s^1,s^2,s^3)^T$.
The direction in which this vector
points in the abstract 3-dimensional
Stokes space uniquely characterizes
the polarization of the radiation.
Stokes vectors lying in the $s^1$-$s^2$
plane correspond to all possible linear
polarizations,
while Stokes vectors parallel and
antiparallel to the $s^3$ axis 
give the two circular polarizations.
General right-handed elliptical
polarizations point in the upper-half Stokes space, $s^3>0$,
while left-handed are given by the lower half, $s^3<0$.

This formalism provides an intuitive
picture of birefringence.
It can be shown that birefringence causes a
rotation of the Stokes vector $\mbf s$
about some axis $\mbf \vs$.
This occurs whenever the usual
degeneracy between the various
polarizations in broken.
Formally,
we solve the modified equations of motion.
We find that some types of violations
lead to two propagating eigenmodes
that have slightly different velocities.
They also differ in polarization,
and light of an arbitrary polarization
is a superposition of the two eigenmodes.
This superposition is altered
as the eigenmodes propagate
at different velocities,
causing an oscillatory effect
that reveals itself as a rotation
of the Stokes vector.
The rotation takes the form
\beq
d\mbf s/dt = 2\om \mbf \vs\times\mbf s \ ,
\label{rot}
\eeq
where $\om$ is the wave frequency,
and the rotation axis $\mbf\vs$
corresponds to the Stokes vector
of the faster eigenmode.
In general, $\mbf\vs$
may depend on both the direction
of propagation and the frequency.

The basic idea of a birefringence
test is to examine light from a
distant polarized source 
for the above rotation.
To do this we need to express the
rotation axis $\mbf \vs$ in terms
of the coefficients for Lorentz violation.
The general result is rather complicated,
but can be written in a relatively simple
form in terms of a set of ``vacuum'' coefficients,
which are linear combinations of the general coefficients.
The calculation involves decomposing $\mbf\vs$ 
into spin-weighted spherical harmonics.
The result takes the form
\begin{align}
\vs^1\mp i\vs^2 &= \sum_{dlm} \om^{d-4} 
(\kE \pm i \kB)\, \sylm{\pm2}{lm}(\mbf{\hat n})\ , \\
\vs^3 &= \sum_{dlm} \om^{d-4} \kV\, \sylm{0}{lm}(\mbf{\hat n})\ ,
\end{align}
where $\sylm{s}{lm}$ is a spin-weighted
spherical harmonic with spin-weight $s$,
and $\mbf{\hat n}$ is the radial unit
vector pointing toward the source on the sky.
The vacuum coefficients
$\kV$, $\kE$, and $\kB$ represent
the minimal combinations 
of coefficients for Lorentz violation
that cause birefringence and affect polarization.
The designations $E$ and $B$ refer to
the parity of the coefficient and
is borrowed from radiation theory.
In the next sections,
we describe a search for these effects
in existing CMB polarization data.

\section{CMB}

The CMB is conventionally
parameterized by a spin-weighted
spherical-harmonic expansion
similar to the expansion of $\mbf\vs$ given above
\cite{cmbrev,polrev}.
The complete characterization
of radiation from a given point
on the sky includes the temperature $T$,
the linear polarization, given by
Stokes parameters $s^1$ and $s^2$,
and the circular polarization, given by $s^3$.
The global description is 
given by the expansion
\begin{gather}
  T=\sum a_{(T)lm}\, \sylm{0}{lm}(\mbf{\hat n})\ ,
  \quad\quad
  s^3=\sum a_{(V)lm}\, \sylm{0}{lm}(\mbf{\hat n})\ , \notag \\
  s^1\mp is^2 = \sum  (a_{(E)lm} \pm i a_{(B)lm})\, \sylm{\pm2}{lm}(\mbf{\hat n})\ . 
\end{gather}
One then constructs various power spectra,
\beq
C^{X_1X_2}_l = \Frac1{2l+1} \sum_m \vev{a^*_{(X_1)lm}a_{(X_2)lm}}\ ,
\eeq
where $X_1,X_2=T,E,B,V$.
These spectra quantify the
angular size variations in
each mode and any correlation 
between different modes.
Smaller $l$ correlates to larger
angular size on the sky.

Within conventional physics,
we expect a nearly isotropic ($l=0$)
temperature distribution.
However, tiny fluctuations in
temperature during recombination
not only introduce higher-order
multipole moments ($l>0$)
but also provide
the necessary anisotropies to
produce a net polarization.
Only linear polarizations are
expected since no circular polarization
is produced in Thomson scattering.
Furthermore, $E$-type polarization
is expected to dominate and be correlated
with the temperature.
No correlation is expected between the
much smaller $B$ polarization 
and temperature.
This general picture agrees with observation
to the extent to which CMB radiation
has been measured
\cite{cmbexp,boompol}.

A breakdown of Lorentz symmetry
may alter these basic features.
Some of the new effects
can be readily understood
as consequences of the Stokes rotations.
For example,
the $CPT$-odd coefficients $\kV$
lead to a Stokes rotation axis that
points along the $s^3$ direction.
The resulting local rotations 
in polarization leave the circularly
polarized component unchanged.
However,
it does lead to a rotation
in the linear components,
causing a simple change in the
polarization angle at each point on
the sky.
Globally this causes a mixing between
the $E$ and $B$ polarization.
This could introduce an unusually
large $B$ component,
which gives a potential signal
of $CPT$ and Lorentz violation.
Similar mixing can arise from
the $\kE$ and $\kB$ coefficients.
However,
since these give a rotation
axis that lies in the $s^1$-$s^2$ plane,
the rotations in this case
also introduce circular polarization.
So a large circularly polarized
component in the CMB might indicate a $CPT$-even 
violation of Lorentz invariance.

All except the $d=3$ coefficients
result in frequency-dependent rotations.
Also, only the $l=0$ coefficients cause
isotropic rotations that are uniform
across the sky.
As a result,
the coefficient $\kVdlm{3}{00}$ 
provides a simple isotropic
frequency-independent special case.
A calculation shows that this
case causes a straightforward
rotation between
$C_l^{EE}$, $C_l^{BB}$, and $C_l^{EB}$,
as well as between
$C_l^{TE}$ and $C_l^{TB}$
\cite{feng}.
In contrast,
more general anisotropic and frequency-dependent
cases cause very complicated mixing
between the various $C_l^{X_1X_2}$
and require numerical integration of
the rotation \rf{rot} over the sky and
frequency range.

\section{Results}

To illustrate the kinds of sensitivities 
that are possible in CMB searches for birefringence,
we next examine the results of the 
BOOMERANG experiment
\cite{boompol}.
This balloon-based experiment
made polarization measurements
in a narrow band of frequencies 
at approximately 145 GHz.
This relatively high frequency
implies that BOOMERANG
is well suited to birefringence tests
since for all violations,
except those with $d=3$,
higher photon energy implies a larger
rotation in polarization.
The small frequency range is also
helpful since we can approximate
all frequencies as $\sim 145$ GHz.

\begin{figure}
  \centerline{\includegraphics[width=\textwidth]{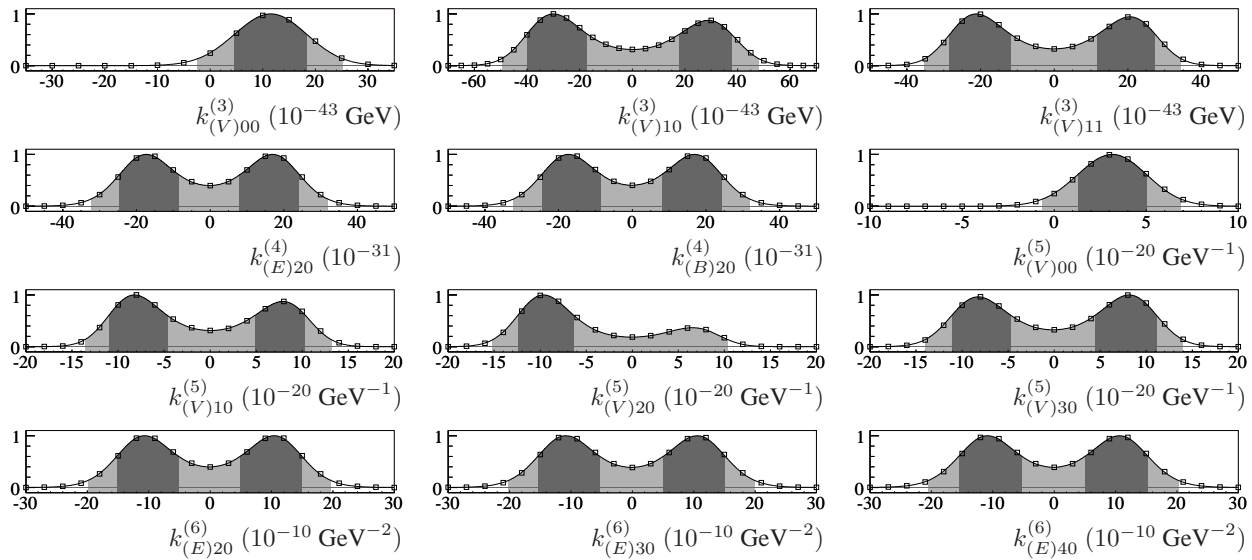}}
  \caption{\label{fig}
    Relative likelihood 
    versus coefficients for Lorentz violation.
    Boxes indicate numerically
    calculated values,
    and the curve is the smooth 
    extrapolation of these points.
    The dark-gray shaded region 
    indicates the 68\% confidence level,
    and the light-gray shows the 95\% level.}
\end{figure}

In our calculation,
we assume conventional polarization
is produced during recombination and
numerically determine the rotated
polarization for points across the sky.
The resulting $C_l^{X_1X_2}$ for various
values of coefficients for Lorentz violation
are determined and compared to published 
BOOMERANG results.
Figure \ref{fig} shows the calculated
relative likelihood
for a sample of 12 coefficients for 
Lorentz violation.
In each case,
we vary the value of one coefficient,
setting all other coefficients to zero.
The $1\si$ and $2\si$ regions are shown.

Some generic features are seen in our survey.
In each case, the coefficient is
nonzero at the $1\si$ level,
hinting at possible Lorentz violation.
However,
since this occurs in every case,
it is likely that this indicates
some systemic feature of the 
BOOMERANG data or our analysis.
We also see that each case
is consistent with no Lorentz violation
at the $2\si$ level,
giving conservative upper
bounds on the 12 coefficients
in Figure \ref{fig}.

These results demonstrate
the potential of the CMB for
testing Lorentz invariance.
Due to the long propagation times,
the sensitivities to $d=3$ coefficients
afforded by the CMB are near the 
limit of what can be expected
in birefringence tests.
However, for $d\geq 4$,
better sensitivities might be
obtained using high-frequency
sources like gamma-ray bursts
\cite{km2}.
Regardless, because of its all-sky nature,
the CMB provides a useful
probe that can simultaneously probe
large portions of coefficient space,
which is difficult in searches involving
a handful of point sources that
access a limited number of propagation directions.

\section*{Acknowledgments}
This work was supported in part
by the Wisconsin Space Grant Consortium.


\begin{thebibliography}{xx}
\def\etal {{\it et al.}}

\bibitem{kost}
  V.A.\ Kosteleck\'y and S.\ Samuel, Phys.\ Rev.\ D {\bf 39}, 683 (1989);
  V.A.\ Kosteleck\'y and R.\ Potting, Nucl.\ Phys.\ B {\bf 359}, 545 (1991).
\bibitem{cpt04}
  For a recent review of various 
  experimental and theoretical
  approaches to Lorentz violation,
  see, for example,
  V.A.\ Kosteleck\'y, ed.,
  {\it CPT and Lorentz Symmetry III, } World Scientific, Singapore, 2005. 
\bibitem{ck} 
  D.\ Colladay and V.A.\ Kosteleck\'y,
  Phys.\ Rev.\ D {\bf 55}, 6760 (1997); {\bf 58}, 116002 (1998);
  V.A.\ Kosteleck\'y, Phys.\ Rev.\ D {\bf 69}, 105009 (2004).
\bibitem{cavities}
  J.\ Lipa \etal, Phys.\ Rev.\ Lett.\ {\bf 90}, 060403 (2003);
  H.\ M\"uller \etal, Phys.\ Rev.\ Lett.\ {\bf 91}, 020401 (2003);
  P.\ Antonini \etal, Phys.\ Rev.\ A {\bf 71}, 050101 (2005);
  S.\ Herrmann \etal, Phys.\ Rev.\ Lett.\ {\bf 95}, 150401 (2005);
  P.L.\ Stanwix \etal, Phys.\ Rev.\ D {\bf 74}, 081101 (2006).
\bibitem{cfj}
  S.M.\ Carroll, G.B.\ Field, and R.\ Jackiw, Phys.\ Rev.\ D {\bf 41}, 1231 (1990).
\bibitem{km1}
  V.A.\ Kosteleck\'y and M.\ Mewes, 
  Phys.\ Rev.\ Lett.\ {\bf 87}, 251304 (2001);
  Phys.\ Rev.\ D {\bf 66}, 056005 (2002).
\bibitem{km2}
  V.A.\ Kosteleck\'y and M.\ Mewes, Phys.\ Rev.\ Lett.\ {\bf 97}, 140401 (2006).
\bibitem{km3}
  V.A.\ Kosteleck\'y and M.\ Mewes, Phys.\ Rev.\ Lett.\ {\bf 99}, 011601 (2007).
\bibitem{photons}
  H.\ M\"uller \etal, Phys. Rev. D {\bf 67}, 056006 (2003);
  R.\ Lehnert and R.\ Potting, Phys.\ Rev.\ Lett.\ {\bf 93}, 110402 (2004);  
  Q.G.\ Bailey and V.A.\ Kosteleck\'y, Phys.\ Rev.\ D {\bf 70}, 076006 (2004);
  M.\ Mewes and A.\ Petroff, Phys.\ Rev.\ D {\bf 75}, 056002 (2007);
  A.J.\ Hariton and R.\ Lehnert, Phys.\ Lett.\ A {\bf 367}, 11 (2007); 
  B.\ Altschul, Phys.\ Rev.\ D {\bf 75}, 105003 (2007);
  A.\ Kobakhidze and B.H.J.\ McKellar, Phys.\ Rev.\ D {\bf 76}, 093004 (2007).
\bibitem{km4}
  V.A.\ Kosteleck\'y and M.\ Mewes, in preparation.
\bibitem{cmbrev}
  For review of current CMB theory and experiment,
  see, for example, 
  Particle Data Group (http://pdg.lbl.gov),
  S.\ Eidelman et al., Phys. Lett. B 592, 1 (2004).
\bibitem{polrev}
  For a review of CMB polarimetry, see, for example,
  W.\ Hu and M.\ White, New Astron.\ {\bf 2}, 323 (1997).
\bibitem{cmbexp}
  G.F.\ Smoot {\it et al.}, Ap.\ J.\ {\bf 396}, L1 (1992);
  J.\ Kovac {\it et al.}, Nature {\bf 420} 772 (2002);
  A.C.S.\ Readnead {\it et al.}, Science {\bf 306}, 836 (2004);
  D.\ Barkats {\it et al.}, Ap.\ J.\ {\bf 619}, L127 (2005);
  E.M.\ Leitch {\it et al.}, Ap.\ J.\ {\bf 624}, 10 (2005);
  W.C.\ Jones \etal, Ap.\ J.\ {\bf 647}, 823 (2006);
  G.\ Hinshaw \etal, Ap.\ J.\ Supp.\ {\bf 170}, 288 (2007);
  L.\ Page \etal, Ap.\ J.\ Suppl.\ {\bf 170}, 335 (2007).
\bibitem{boompol}
  T.E.\ Montroy \etal, Ap.\ J.\ {\bf 647}, 813 (2006);
  F.\ Piacentini \etal,  Ap.\ J.\ {\bf 647}, 833 (2006).
\bibitem{feng}
  B.\ Feng {\it et al.}, Phys.\ Rev.\ Lett.\ {\bf 96}, 221302 (2006);
  P.\ Cabella \etal, Phys.\ Rev.\ D {\bf 76}, 123014 (2007).
\end{thebibliography}
\end{document}